\documentclass[fleqn]{annalen} 
\usepackage{graphics} 
\begin{document} 
\newcommand{\volume}{8}              
\newcommand{\xyear}{1999}            
\newcommand{\issue}{5}               
\newcommand{\recdate}{29 July 1999}  
\newcommand{\revdate}{dd.mm.yyyy}    
\newcommand{\revnum}{0}              
\newcommand{\accdate}{dd.mm.yyyy}    
\newcommand{\coeditor}{ue}           
\newcommand{\firstpage}{1}           
\newcommand{\lastpage}{4}            
\newcommand{\beq}{\begin{equation}} 
\newcommand{\eeq}{\end{equation}} 
\newcommand{\bea}{\begin{eqnarray}} 
\newcommand{\eea}{\end{eqnarray}} 
\setcounter{page}{\firstpage}        
\newcommand{\keywords}{bifurcation, quantum chaos, localized states}  
\newcommand{\PACS}{03.65.Sq, 05.45.Mt, 72.15.Rn} 
\newcommand{\shorttitle}{I. Varga et al., Localized states near bifurcations}
\title{Quantum Localization near Bifurcations in Classically Chaotic 
  Systems} 
\author{I.\ Varga, P.\ Pollner, and B.\ Eckhardt}  
\newcommand{\address} 
  {Fachbereich Physik, Philipps--Universit\"at Marburg, 
  D-35032 Marburg, Germany  
} 
\newcommand{\email}{\tt varga@phy.bme.hu}  
\maketitle 
\begin{abstract} 
We show that strongly localized wave functions occur around classical 
bifurcations. Near a saddle node bifurcation the scaling of the inverse 
participation ratio on Planck's constant and the dependence on the parameter 
is governed by an Airy function. Analytical estimates are supported by 
numerical calculations for the quantum kicked rotor. 
\end{abstract} 
 
\medskip
Many mechanisms of localization in dynamical systems 
have been discussed. The saturation in the spreading 
of a wave packet in the quantum kicked rotor can 
be connected to Anderson localization \cite{Casati79, GPF84}. 
Other effects, such as trapping due to cantori \cite{Geisel86, Bruno99} 
or in elliptic islands, have a classical origin. 
The scarring phenomenon, an increase in intensity
near classical periodic orbits, seems to be intermediate
\cite{Heller, Kaplan}. Scars are most prominent
on orbits that are weakly unstable or marginally stable,
as in the stadium billiard. This suggests that near
a classical bifurcation, where the semiclassical amplitudes
diverge, they should be most prominent. As we will show
here this is indeed the case: the classical slowing down 
in phase space near a bifurcation causes an enhanced eigenfunction 
amplitude, but there is also a quantum interference between two 
orbits that brings in quantum modulations. Specifically, we here 
study the localization of the wave functions in the quantized standard 
map near saddle node bifurcations of period one orbits. 
 
The quantum kicked rotor is described by a finite unitary matrix 
which in the momentum plane wave representation is given by 
\cite{Izr90} 
\beq 
   U_{nm}=\frac{1}{N}\sum_{l=0}^{N-1} 
          e^{-ik\cos(\frac{2\pi l}{N})} 
          e^{-i\frac{2\pi}{N}l(n-m)}e^{-i\frac{\tau}{2}m^2} \,. 
\label{umatrix} 
\eeq 
The dimension $N$ of the momentum basis is related to Planck's 
constant by $\hbar=4\pi/N$. The parameters $k$ and $\tau$ 
in (\ref{umatrix}) are connected to $\hbar$ and the classical 
kicking strength parameter, $K$, by $k=K/\tau$ and $\tau=4\pi/N$, 
where the latter choice ensures the absence of dynamical localization. 
In the classical limit the quantum kicked rotor corresponds to the 
standard map defined on a torus,
\beq
   p_{n+1}=p_n+K\sin\,\vartheta_n \mbox{\ mod\ }4\pi, \qquad\qquad  
   \vartheta_{n+1}=\vartheta_n+p_{n+1} \mbox{\ mod\ } 2\pi\,.
\eeq 
The classical standard map depends on $K$ only. For $K>0$ the 
phase space is mixed and at $K\approx 0.93996...$ the last KAM torus 
breaks up. For $K>5$ the system is essentially fully mixing with 
exponentially small stable islands. However, even for sufficiently 
large $K$ values the classical phase space is very complex. 
Of particular interest here are parameter values near a 
saddle node bifurcation where pairs of periodic orbits 
are created or destroyed. Such bifurcations carry large 
semiclassical weights since the second derivative of the  
action vanishes and the semiclassical amplitude diverges. 
The effect of such bifurcations on the spectral statistics 
has been much studied~\cite{Ber98}. Here we focus on the wave functions.  
 
\begin{figure}[tbh] 
\centerline{\resizebox{12.25cm}{4.25cm}{\includegraphics{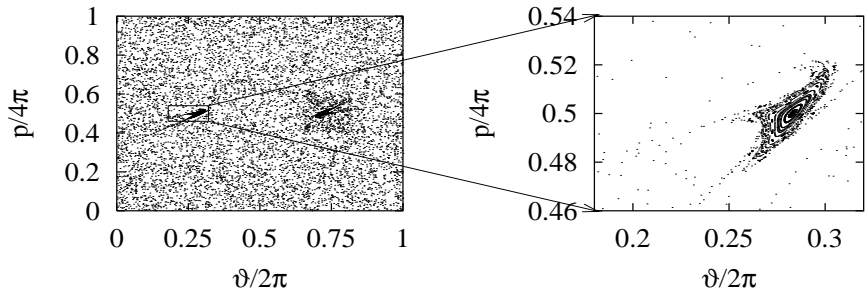}}} 
   \caption{Classical phase space near the bifurcation for $K/4\pi =1.0216$.
   The bifurcation takes place at ($\vartheta_0=\pi/2$, $p_0=2\pi$)} 
\label{phspace} 
\end{figure}

The simplest examples of saddle node bifurcations arise when 
$K$ passes through $K^*=4\pi$. The two pairs of  
stable and unstable period one orbits are created.  
Fig.~\ref{phspace} shows the phase space structure for a $K$-value 
slightly above the bifurcation, $K/4\pi=1.0216$. 
 
The quantity we use to detect localization in wave functions is the 
average inverse participation ratio of the wave functions, i.e. the fourth 
moment of their components $c_{\alpha i}=\langle i|\alpha\rangle$ of the 
$\alpha$'s eigenstate in some orthonormal basis $|i\rangle$ summed over all 
eigenstates and normalized by the value expected within random matrix 
theory for the orthogonal ensemble \cite{IV95,Guhr98}, 
\beq 
   L=\frac{1}{3}\sum_{\alpha}\sum_i\,|c_{\alpha i}|^4\,. 
\label{locmeas} 
\eeq 
Thus, if all wave functions are ergodically spread over all basis states, 
$L=1$. Localized states will show an $L>1$. This is a global quantity 
that is related to the return probability which can be calculated in 
the following way. The transition probability from state $i$ to state $j$ 
after $m$ kicks is $P_{ij}(m)=|\mbox{Tr}(A_{ij}U^m)|^2$, 
where $A_{ij}$ is a projector $A_{ij}=|j\rangle\langle i|$ from the $i$--th 
to the $j$--th basis state. The localization measure $L$ is nothing
but the average long time return probability, $L=1/3\sum_i {\cal P}_{ii}$, 
where ${\cal P}_{ii}=\lim_{M\to\infty}\sum_{m=1}^M P_{ii}(m)$ is the
return probability to state $|i\rangle $.

The quantity $L$ can also be estimated using the semiclassical procedures  
proposed in~\cite{Bruno95} for matrix elements. Specifically, let  
$A_{ii}=|i\rangle\langle i|$ be the projector on the $i$-th basis state. Then 
$\rho_{A_{ii}}=\sum_{\alpha}|\langle i|\alpha\rangle |^2  
\delta (\varphi-\varphi_{\alpha})$ is the matrix element weighted density  
of states. The semiclassical expression for $\rho_{A_{ii}}$ has
two terms, $\rho_{A_{ii}} = \rho_0(A_{ii})+ \rho_{osc}$, where the first 
term is the smooth phase space average and comes from the Thomas-Fermi 
density of states, whereas the latter part can be expressed semiclassically
as a sum over periodic orbits. The weight of each periodic orbit $p$ also 
depends on the integral of the observable along the points of the orbit. 
\begin{figure}[tbh] 
\centerline{\resizebox{4.25cm}{4.25cm}{\includegraphics{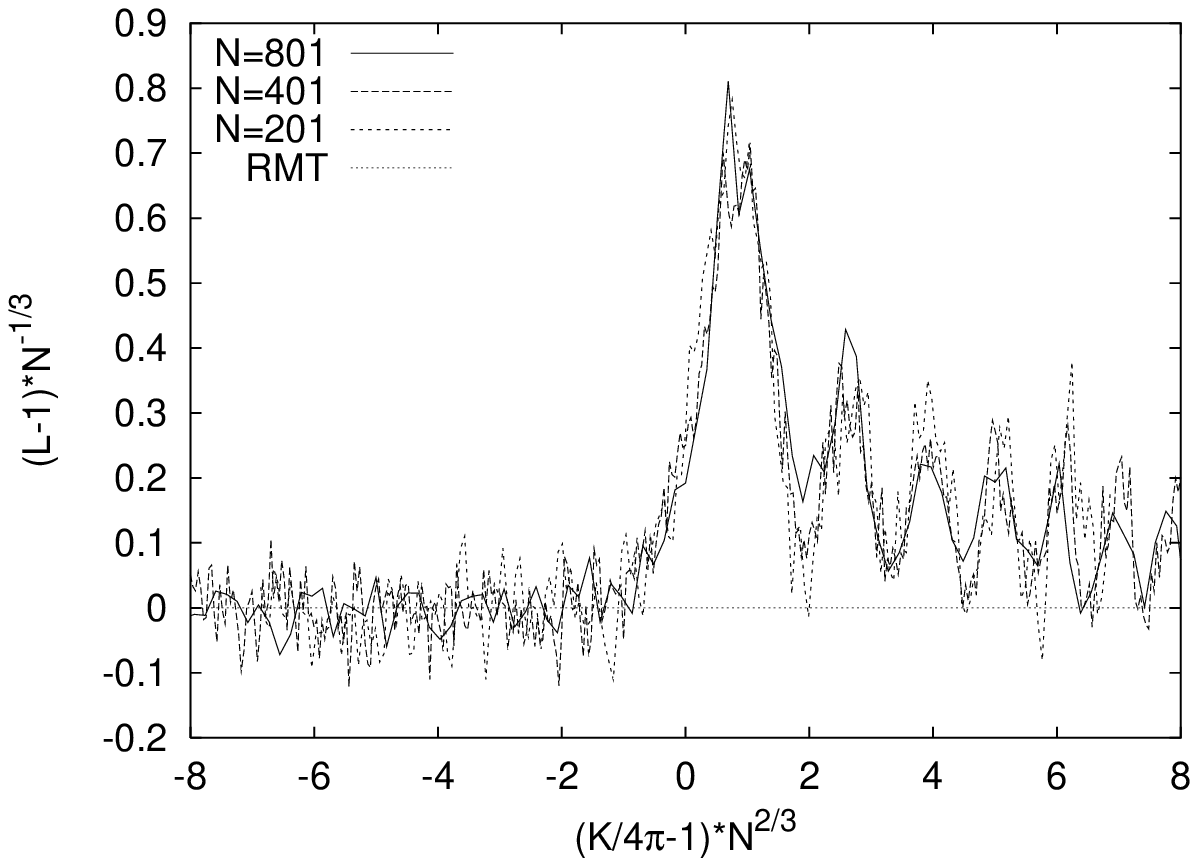}}} 
   \caption{Measure of localization in the vicinity of the saddle--node 
   bifurcation at $K/4\pi=1$ for different values of $N=201$, 401 and 
   801. The curves have been rescaled according to the 
   uniformization expected from Eq.~(\protect\ref{loc-sc}).}
\label{partr-locst} 
\end{figure} 
The fourth moment that enters the expression for $L$ is a measure of 
the fluctuations of the matrix elements $\langle\alpha 
|A_{ii}|\alpha\rangle=|\langle i|\alpha\rangle|^2=|c_{\alpha i}|^2$ 
around their average and can hence be calculated from the square of 
the sum over periodic orbits. Within the diagonal approximation we 
therefore obtain 
\beq 
   \left\langle |c_{\alpha i}|^4\right\rangle_{\alpha}- 
   \left\langle |c_{\alpha i}|^2\right\rangle_{\alpha}^2 
   \simeq\sum_p |w_p|^2 |A_p^{(i)}|^2 \, 
\label{vari} 
\eeq 
where the $w_p$ contain the stability information about the orbit
and the $A_p^{(n)}$ the contribution from the observable 
\cite{Bruno95,Bruno95b}. When all orbits are uniformly hyperbolic the 
random matrix value is recovered. However, in our situation near a 
bifurcation we get diverging amplitudes $w_p$ which, when expressed 
in a uniform approximation, are replaced by an Airy function. For the 
fixed points near $K^{\ast}/4\pi=$integer, we can write 
$\rho_{osc}=\sum_p' t_p + t_{bif}$, where the sum extends over  
all orbits except the ones near the bifurcation. The weights $t_p$ are 
given in a conventional form~\cite{Bruno95} whereas the bifurcating 
orbit contribution~\cite{pb} is 
$t_{bif}\approx\tilde{A}_{bif}N^{1/6}\mbox{Ai}[N^{2/3}(K-K^{\ast})]$. 
Therefore the fluctuations are expected to show a contribution which 
is a square of an Airy function  
\beq 
   L-1\simeq N^{1/3}\mbox{Ai}^2[bN^{2/3}(K-K^{\ast})]. 
\label{loc-sc} 
\eeq 
 
In Fig.~\ref{partr-locst} we present the variation of $L$ as a function 
of $K$ around the saddle--node bifurcation of period--one orbits 
at $K^{\ast}/4\pi=1$ for different values of $N$. For $K<K^{\ast}$ the 
random matrix behavior is recovered with increasing $N$. Beyond $K^{\ast}$ 
Airy--type oscillations appear. Both the $K$ and the $L$ 
values have been rescaled according to the $N$--scaling in (\ref{loc-sc}). 
After rescaling the first peaks for different $N$ the curves fall on top of 
each other. Higher than second moments of the eigenvector components  
show an even enhanced effect provided the distribution  
function of $|c_{\alpha i}|^2$ develops a tail due to the presence of 
localized states.
 
\begin{figure}[tbh] 
\centerline{\resizebox{4.25cm}{4.25cm}{\includegraphics{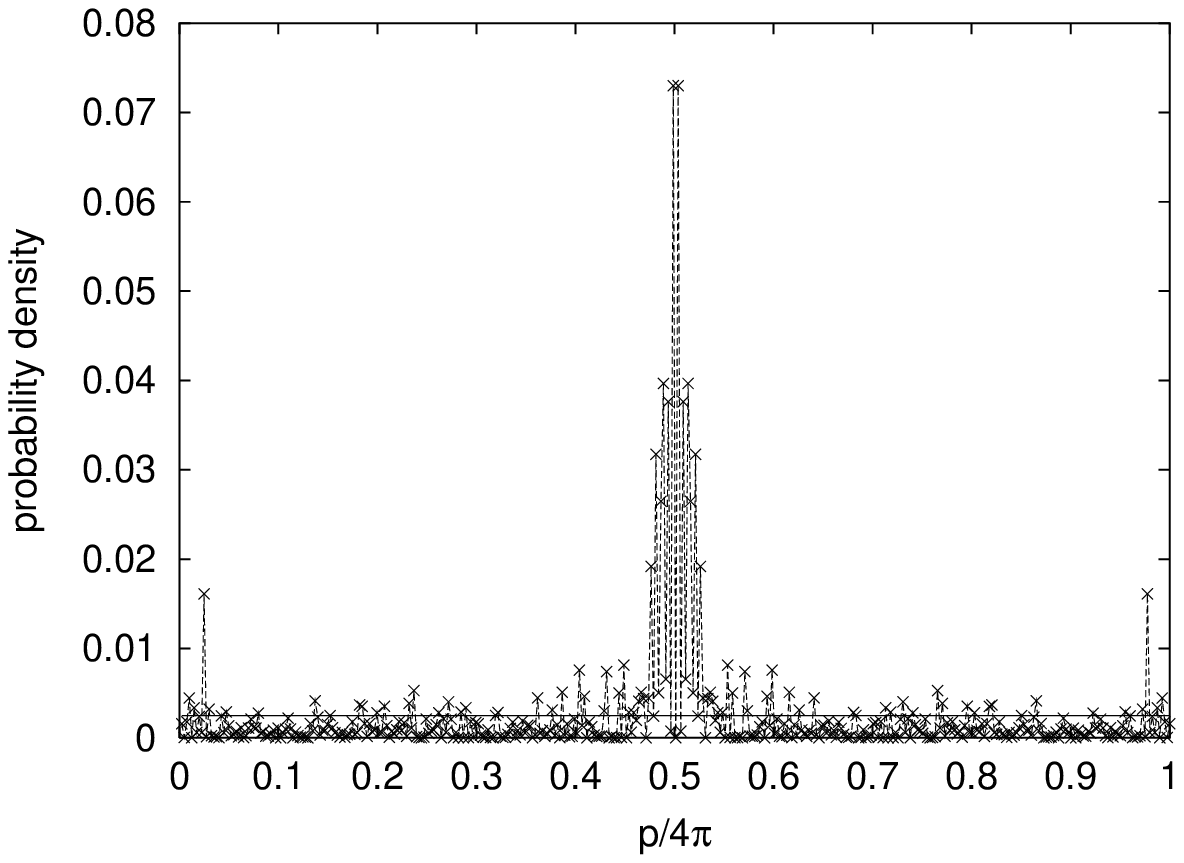}} 
            \resizebox{4.25cm}{4.25cm}{\includegraphics{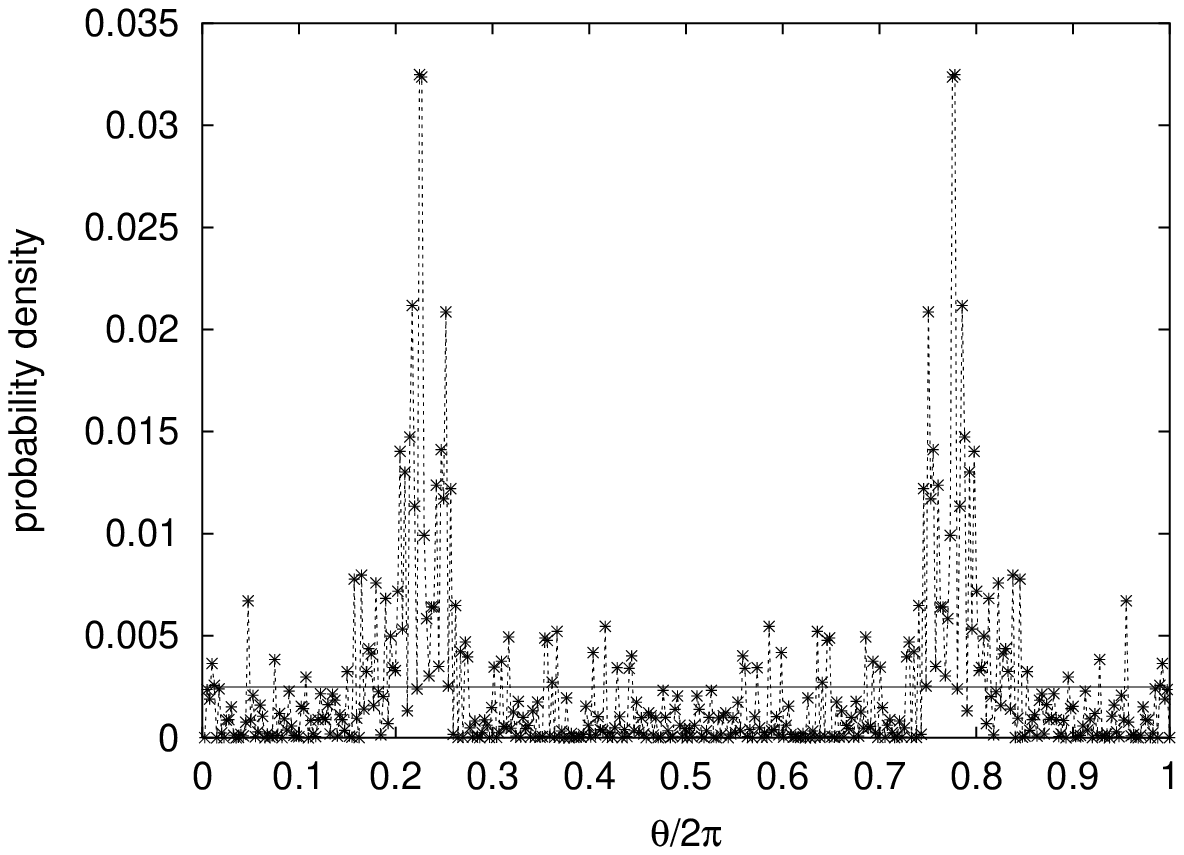}}} 
   \caption{A localized state at $K=1.0216$ for $N=401$ in $p-$ (left panel) 
   and $\vartheta-$representation (right panel). The constant line indicates 
   the mean value expected from random matrix theory.} 
\label{locstate} 
\end{figure}

As an illustration in Fig.~\ref{locstate} we show the state with the 
largest inverse participation ratio at the value of $K/4\pi=1.0216$ for 
$N=401$. This state is localized at $p=2\pi$. In $\vartheta$-representation 
the state is indeed localized around $\vartheta=\pi/2$ and 
$\vartheta=3\pi/2$, right at the pair of orbits that is born in the  
saddle--node bifurcation. Due to the uncertainity principle 
its spatial extent is larger than in $p$-representation. 
 
In summary, we have analyzed the localization of quantum eigenstates 
in the vicinity of bifurcating orbits. The scaling and uniformization 
predictions based on the semiclassical theory describe the variations of 
global quantities like the average inverse participation number of the states. 
We would like to stress that the properties of the fixed points 
(e.g. through local Lyapunov exponent as in Hellers scar theory) 
do not explain the localization since the effect of the bifurcation is 
seen over a larger scale in the parameter $K$ than expected from the 
appearance of the stable island and the local escape time calculated 
from the Lyapunov exponent. Such a theory could also not explain the 
interference pattern seen in the inverse participation ratio.
 
\vspace*{0.25cm} \baselineskip=10pt{\small \noindent Financial support  
of the Alexander von Humboldt Foundation, the KAAD and the OTKA Nos. 
F024135, T024136, T029813 are gratefully acknowledged.} 

\end{document}